\documentclass[]{interact}

\usepackage{epstopdf}
\usepackage[caption=false]{subfig}

\usepackage[numbers,sort&compress,merge]{natbib}
\bibpunct[, ]{[}{]}{,}{n}{,}{,}
\renewcommand\bibfont{\fontsize{10}{12}\selectfont}

\theoremstyle{plain}

\theoremstyle{definition}

\theoremstyle{remark}

\usepackage{enumitem}

\newcommand{\Tr}{\operatorname{Tr}}

\newcommand{\Eg}{E_\mathrm{g}}
\newcommand{\Ee}{E_\mathrm{e}}
\newcommand{\rgg}{\rho_\mathrm{gg}}
\newcommand{\rge}{\rho_\mathrm{ge}}
\newcommand{\reg}{\rho_\mathrm{eg}}
\newcommand{\ree}{\rho_\mathrm{ee}}

\begin{document}

\articletype{Paper for Special Issue ``Quantum Dynamics and Resonances in Chemistry and Physics"}

\title{Exceptional points of the Lindblad operator of a two-level system}

\author{
\name{Naomichi Hatano\thanks{CONTACT Naomichi Hatano. Email: hatano@iis.u-tokyo.ac.jp}}
\affil{Institute of Industrial Science, The University of Tokyo\\
5-1-5 Kashiwanoha, Kashiwa, Chiba 277-8574, Japan}
}

\maketitle

\begin{abstract}
The Lindblad equation for a two-level system under an electric field is analyzed by mapping to a linear equation with a non-Hermitian matrix.
Exceptional points of the matrix are found to be extensive;
the second-order ones are located on lines in a two-dimensional parameter space, while the third-order one is at a point.
\end{abstract}


\section{Introduction: non-Hermitian quantum mechanics}
\label{sec1}
Non-Hermitian quantum mechanics~\cite{Moiseyev11} is attracting evermore interest from various points of view.
%
Historically, the research of non-Hermitian quantum mechanics was initiated in the field of nuclear physics;
we can go back to Gamow's study in 1928~\cite{Gamow28} to find an explanation of resonant scattering in terms of a resonant state with a complex eigenvalue.
The resonant state was then defined as an eigenstate of the Schr\"{o}dinger equation under the Siegert boundary condition~\cite{Siegert39}, which is essentially a non-Hermitian condition, and hence can produce the complex eigenvalue.

In 1950s there appear many studies on ``the optical model''~\cite{Feshbach58review} to explain the nuclear decay phenomenologically.
Feshbach~\cite{Feshbach58,Rotter09} justified the phenomenology by showing that an effective complex potential comes out when one traces out the environmental space surrounding the central scattering area.
We can understand this in terms of the equivalence~\cite{Hatano13} of the Feshbach formalism to the Siegert boundary condition; the non-Hermiticity implicitly hidden in the Siegert boundary condition for the outer space emerges explicitly when one traces out the environment.

This field of research in fact can be called the work on open quantum systems~\cite{Hatano14} in the modern terminology.
Many studies on non-Hermitian random matrices (\textit{e.g.}\ Ref.~\cite{Feinberg97}) were motivated by the non-Hermiticity that appears in this way.
We should also note that the non-Hermiticity assumed \textit {a priori} in many recent studies on topological aspects of non-Hermitian systems (\textit{e.g.}\ Refs.~\cite{Kunst18,Yao18,Yoshida18})
is implicitly based on the work on open quantum systems.
Experiments on non-Hermitian quantum systems (\textit{e.g.}\ Ref.~\cite{Xiao17}) also realize the non-Hermiticity by means of the contact to the environment.

%
%
%

One of the interesting phenomena that are specific to non-Hermitian systems including such open quantum systems is the exceptional point.
In Hermitian systems, when two eigenvalues become equal to each other, the Hamiltonian in the subspace of the eigenvalues is diagonalized into a unit matrix times the degenerate eigenvalue as in
\begin{align}
H=\begin{pmatrix}
\lambda_\mathrm{deg} & 0 \\
0 & \lambda_\mathrm{deg}
\end{pmatrix},
\end{align}
and the eigenvectors of the two eigenvalues are intact: 
\begin{align}\label{eq2}
\begin{pmatrix}
1 \\
0
\end{pmatrix}
\quad
\mbox{and}
\quad
\begin{pmatrix}
0\\
1
\end{pmatrix}.
\end{align}
In non-Hermitian systems, on the other hand, when two eigenvalues coalesce, the Hamiltonian in the subspace cannot be diagonalized but can be transformed only to the Jordan block:
\begin{align}
H=\begin{pmatrix}
\lambda_\mathrm{coal} & 1 \\
0 & \lambda_\mathrm{coal}
\end{pmatrix}.
\end{align}
The eigenvectors coalesce too, and only the first of the two in Eq.~\eqref{eq2}  survives.
The point at which the coalescence takes place is referred to as an exceptional point~\cite{Kato95,Heiss12}.

More specifically, the above is the second-order exceptional point; 
it is called the $p$th-order one when $p$ pieces of eigenvalues coalesce.
When one moves around a $p$-th order exceptional point adiabatically, the state alternates from one eigenstate to another out of the $p$ pieces eigenstates that coalesce.
It has been reported that the state rather converges to one of the eigenstates when one moves around the exceptional point \textit{non}-adiabatically~\cite{Uzdin11,Gilary13,Doppler16}.

In mathematics, it is often termed that non-Hermitian matrices are ``generally \textit{non}-diagonalizable.''
In practice, however, non-Hermitian matrices are generally diagonalizable \textit{except} at the exceptional points, by which we mean that the exceptional points typically exist only at separate points.
Indeed, consider a general case of the coalescence of two complex eigenvalues.
When we equate the eigenvalues to obtain the position of a second-order exceptional point, we generally have two conditions for each of the real parts and the imaginary parts.
This therefore fixes two system parameters. 
In other words, the exceptional point generally appears as discrete points in a two-dimensional parameter space.
Argument on the same line dictates that a third-order exceptional point emerges only in a four-dimensional parameter space.

In the present study, in contrast, we show an incidence in which the second-order exceptional ``point'' forms two continuous curves in a two-dimensional parameter space and a third-order exceptional point appears as a meeting endpoint of the two curves.
This is due to a symmetry that the system observes;
the imaginary parts of two eigenvalues to coalesce are always equal to each other, and therefore the condition for the coalescence comes from equating only their real parts.
The system is simple enough to be realizable in experiments as will be presented in the next section.

\section{Lindblad equation for a two-level system}

\begin{figure}
\centering
\includegraphics[width=0.3\textwidth]{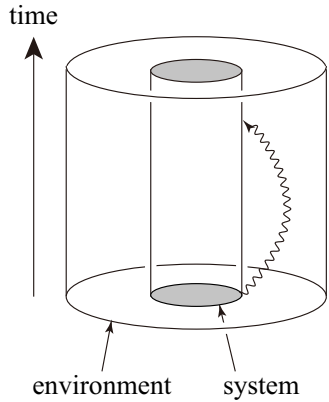}
\caption{A schematic illustration of the time evolution of a system in question connected to the environment.
Perturbation that originates in the system can travel through the environmental degrees of freedom as the wavy arrow indicates, and affects the system at a later time, resulting in a seemingly non-Markovian time evolution of the system.}
\label{fig1}
\end{figure}

We consider a quantum Markovian dynamics of a two-level system in a dissipative environment subjected to an oscillatory electric field.
Consider the von Neumann equation for a density matrix of the entire system that contains the system in question and the environment:
\begin{align}
i\frac{d}{dt}\rho_\mathrm{tot}=\left[H_\mathrm{tot},\rho_\mathrm{tot}\right],
\end{align}
where we put $\hbar$ to unity.
This is a Markovian equation of motion.
Upon tracing out the environmental degrees of freedom, however, the density matrix of the system in question evolves in time generally in a non-Markovian dynamics.
This is because perturbation that originates in the system in question at one point of the time can propagate through the hidden degrees of freedom of the environment and come back to the system at a much later time, which would seem non-Markovian from the point of view of the time evolution of the system alone (Fig.~\ref{fig1}).

It is, however, generally quite difficult to trace out the environment exactly.
Most studies of open quantum systems are done under Markovian approximation; only occasionally the non-Markovianity has been taken into account perturbatively (\textit{e.g.}\ Ref.~\cite{Guarnieri16}).
The most general Markovian approximation of the system is given by the Lindblad equation~\cite{Lindblad76,Gorini76}:
\begin{align}\label{eq10}
i\frac{d\rho}{dt}=[H,\rho]+i\Gamma\left(c\rho c^\dag-\frac{1}{2}c^\dag c\rho-\frac{1}{2}\rho c^\dag c\right),
\end{align}
where $\Gamma$ represents the effect of the coupling to the environment in this type of Markovian approximation.
The Markovian approximation is generally valid when the coupling $\Gamma$ is small compared to a typical energy scale of the system $H$, as can be easily imaginable from the argument above.

The coupling $\Gamma$ gives rise to dissipation to the environment.
As we will see below, it algebraically generates the non-Hermiticity in the time-evolution matrix, and thereby the imaginary parts of its eigenvalues.
More specifically, there are three eigenvalues with negative imaginary parts in addition to a null eigenvalue.
The eigenstates of the former eigenvalues decay exponentially in time owing to the negative imaginary parts and only the eigenstate of the null eigenvalue survives.
This physically means that the system loses the three eigenstates because of the energy dissipation to the environment and converges to the null eigenstate, which corresponds to the equilibrium state.

\begin{figure}
\centering
\includegraphics[width=0.5\textwidth]{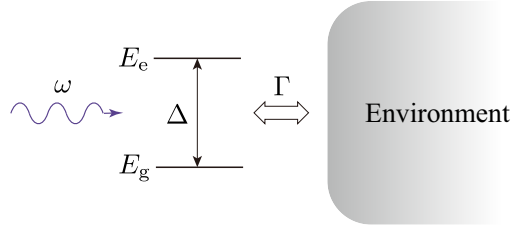}
\caption{We consider a two-level system coupled to an environment with strength $\Gamma$ and under an electric field of amplitude $E_0$ and frequency $\omega$.}
\label{fig2}
\end{figure}

We hereafter consider the Lindblad dynamics of a two-level system under an electric field $\mathcal{E}(t)=E_0\cos(\omega t)$ with a dissipation $\Gamma$ to an environment; see Fig.~\ref{fig2}.
The starting Hamiltonian of the system is 
\begin{align}
\begin{array}{rrl}
&
& 
\begin{array}{cc}
\langle\mathrm{e}|\quad & \quad\langle\mathrm{g}|\quad
\end{array}
\\
H=\hspace{-2ex}
&
\begin{array}{r}
|\mathrm{e}\rangle\\
|\mathrm{g}\rangle
\end{array}
&
\hspace{-2.5ex}
\begin{pmatrix}
\Ee -\Eg & D\mathcal{E}(t) \\
D^\ast\mathcal{E}(t) & 0
\end{pmatrix},
\end{array}
\end{align}
where $\Eg$ and $\Ee$ respectively denote the energy of the ground state $|\mathrm{g}\rangle$ and the excited state $|\mathrm{e}\rangle$, while $D$ denotes the electric dipole moment.
After the rotating-wave approximation, in which we drop one component of the electric-field oscillation $E_0\cos(\omega t)$, we have
\begin{align}\label{eq30}
H&=\begin{pmatrix}
\Delta & de^{-i\omega t}/2 \\
d e^{i\omega t}/2 & 0
\end{pmatrix},
\end{align}
where
\begin{align}
\Delta = \Ee-\Eg
\end{align}
is the level difference and
\begin{align}
d=DE_0
\end{align}
is the effective amplitude of the electric field,
which we assumed here is real.

For the operators $c^\dag$ and $c$ in Eq.~\eqref{eq10}, we assume 
\begin{align}\label{eq60}
c^\dag=\begin{pmatrix}
0 & 1 \\
0 & 0
\end{pmatrix},
\qquad
c&=\begin{pmatrix}
0 & 0 \\
1 & 0
\end{pmatrix}.
\end{align}
The former describes the excitation of the system due to perturbation from the environment, while the latter the relaxation due to energy dissipation to the environment.
In order for the dynamics to be unitary, the coupling parameter $\Gamma$ should be real.

We can get rid of the time dependence of the Hamiltonian~\eqref{eq30} using the unitary transformation
\begin{align}
U(t)=\begin{pmatrix}
e^{-i\omega t} & 0 \\
0 & 1
\end{pmatrix},
\end{align}
which adds a time-dependent phase factor only to the excited level.
Details of the algebra are given in Appendix~\ref{appB}.
We thereby arrive at the updated Lindblad equation of the same form as Eq.~\eqref{eq10},
but with the Hamiltonian now given by a time-independent one,
\begin{align}\label{eq110}
H=\begin{pmatrix}
\Delta-\omega & d/2 \\
d/2 & 0
\end{pmatrix}
=
\begin{pmatrix}
\delta & d/2 \\
d/2 & 0
\end{pmatrix}
\end{align}
instead of the time-dependent rotating-wave approximation~\eqref{eq30}, where
\begin{align}
\delta=\Delta-\omega=\Ee-\Eg-\omega
\end{align}
is the parameter that represents the detuning of the electric field.

We now notice that there are three parameters $\Gamma$, $d$ and $\delta$, but one of them should provide the unit of the energy, and therefore there are essentially only two free parameters.
We hereafter adopt the detuning parameter $\delta$ as the unit of the energy and explore a phase diagram in the parameter space of $d/\delta$ and $\Gamma/\delta$.

\section{Matrix representation of the Lindblad operator}

Let us use the notation
\begin{align}\label{eq40}
\rho(t)&=\begin{pmatrix}
\ree(t) & \reg(t) \\
\rge(t) & \rgg(t)
\end{pmatrix}
\end{align}
for the density matrix in Eq.~\eqref{eq10} and find the time evolution of each of the four elements.
We explicitly write down the Lindblad equation~\eqref{eq10} using the expressions~\eqref{eq60},~\eqref{eq110} and~\eqref{eq40}.
After a straightforward algebra, we find the equations for the four elements as follows:
\begin{align}\label{eq150}
i\dot{\rho}_\mathrm{eg}&= \delta\reg-\frac{d}{2}(\ree-\rgg)-\frac{i}{2}\Gamma\reg,
\\
i\dot{\rho}_\mathrm{ge}&= -\delta\rge+\frac{d}{2}(\ree-\rgg)-\frac{i}{2}\Gamma\rge,
\\
i\dot{\rho}_\mathrm{ee}&= -\frac{d}{2}(\reg-\rge) -i\Gamma\ree,
\\\label{eq180}
i\dot{\rho}_\mathrm{gg}&= \frac{d}{2}(\reg-\rge) +i\Gamma\ree.
\end{align}
Note that the Hermiticity of the density matrix~\eqref{eq40} dictates the following symmetries:
\begin{align}
\reg^\ast=\rge, \quad
\ree^\ast=\ree, \quad
\mbox{and}
\quad
\rgg^\ast=\rgg,
\end{align}
which are indeed observed by Eqs.~\eqref{eq150}--\eqref{eq180}.
By using the Hilbert-Schmidt expression of the density matrix
\begin{align}
\Psi(t)=\begin{pmatrix}
\reg(t) \\
\rge(t) \\
\ree(t) \\
\rgg(t)
\end{pmatrix},
\end{align}
we can cast the Lindblad equation~\eqref{eq10} into the form of a linear equation,
\begin{align}\label{eq190}
i\frac{d}{dt}\Psi(t)=L\Psi(t),
\end{align}
where the Lindblad super-operator $L$ is  given by
\begin{align}\label{eq200}
L=\begin{pmatrix}
\delta-i\Gamma/2 & 0 & -d/2 & d/2 \\
0 & -\delta-i\Gamma/2 & d/2 & -d/2 \\
-d/2 & d/2 & -i\Gamma & 0 \\
d/2 & -d/2 & i\Gamma & 0
\end{pmatrix}.
\end{align}
For $\Gamma=0$, namely without the dissipation, we have only real eigenvalues $0$, $0$, and $\pm\sqrt{\delta^2+d^2}$.
Complex eigenvalues emerge when we turn on the dissipation as in $\Gamma>0$, as we will see below.


Before showing the eigenvalue distributions explicitly, let us here point out a symmetry that the matrix~\eqref{eq200} observes.
Because the Hamiltonian~\eqref{eq110} and the operators~\eqref{eq60} are all real matrices, we have
\begin{align}
-L^\ast=\begin{pmatrix}
-\delta-i\Gamma/2 & 0 & d/2 & -d/2 \\
0 & \delta-i\Gamma/2 & -d/2 & d/2 \\
d/2 & -d/2 & -i\Gamma & 0 \\
-d/2 & d/2 & i\Gamma & 0
\end{pmatrix}.
\end{align}
This becomes identical to $L$ after exchanging the first and second rows as well as the first and second columns.
Therefore, the two matrices $L$ and $-L^\ast$ share the eigenvalues.
Because of this symmetry, the eigenvalue distribution must be symmetric with respect to the imaginary axis, which we will demonstrate below.
This is the important point that gives rise to the fact that the second-order exceptional points form curves in the two-dimensional parameter space of $d/\delta$ and $\Gamma/\delta$.

The eigenvalues of the matrix $L$ are given as follows.
First, the matrix always have a null eigenvalue $z_0=0$ with the left- and right-eigenvectors
\begin{align}
\Phi^L_0=
\begin{pmatrix}
0 &
0 &
1 &
1 
\end{pmatrix},
\qquad
\Phi^R_0=\frac{1}{N_0}
\begin{pmatrix}
-d\left(2\delta+i\Gamma\right) \\
-d\left(2\delta-i\Gamma\right) \\
d^2 \\
4\delta^2+d^2+\Gamma^2 
\end{pmatrix},
\end{align}
which corresponds to the equilibrium state.
The normalization constant $N_0$ is fixed by $\Phi^L_0\Phi^R_0=1$, which is equivalent to $\Tr\rho=\ree+\rgg=1$. 
The equilibrium state is therefore given by
\begin{align}
\rho^\mathrm{eq}=\frac{1}{4\delta^2+2d^2+\Gamma^2}
\begin{pmatrix}
d^2 & -d\left(2\delta+i\Gamma\right) \\
-d\left(2\delta-i\Gamma\right) &4\delta^2+d^2+\Gamma^2 
\end{pmatrix},
\end{align}
which is indeed a Hermitian matrix.

All other three eigenvalues have negative imaginary parts, never being degenerate with the null eigenvalue for real and nonzero values of the parameters, and hence decay in time, giving way to the equilibrium state. 
They are specifically given by
\begin{align}\label{eq220}
z_1&=-i\left(\frac{2}{3}\Gamma+u+v\right),
\\\label{eq230}
z_2&=-i\left(\frac{2}{3}\Gamma+e^{2\pi i/3}u +e^{-2\pi i/3}v\right),
\\\label{eq240}
z_3&=-i\left(\frac{2}{3}\Gamma+e^{-2\pi i/3}u +e^{2\pi i/3}v\right),
\end{align}
where
\begin{align}\label{eq250}
u=\sqrt[3]{q+\sqrt{p^3+q^2}}
\quad\mbox{and}\quad
v=\sqrt[3]{q-\sqrt{p^3+q^2}}
\end{align}
with
\begin{align}
p=\frac{1}{3}\left(\delta^2+d^2-\frac{\Gamma^2}{12}\right)
\quad\mbox{and}\quad
q=\frac{\Gamma}{6}\left(\delta^2-\frac{d^2}{2}+\frac{\Gamma^2}{36}\right).
\end{align}
The corresponding eigenvectors are respectively given by
\begin{align}
\Phi^L_\nu&=
\begin{pmatrix}
2z_\nu\left[(i\Gamma+z_\nu)(i\Gamma+2\delta+2z_\nu)-d^2\right]\\
-2d^2 z_\nu \\
-d(-i\Gamma+z_\nu)(i\Gamma+2\delta+2z_\nu) \\
d(i\Gamma+z_\nu)(i\Gamma+2\delta+2z_\nu) 
\end{pmatrix}^T,
\\
\Phi^R_\nu&=\frac{1}{N_\nu}\begin{pmatrix}
2\left[(i\Gamma+z_\nu)(i\Gamma+2\delta+2z_\nu)-d^2\right] \\
-2d^2 \\
-d(i\Gamma+2\delta+2z_\nu) \\
d(i\Gamma+2\delta+2z_\nu) 
\end{pmatrix},
\end{align}
where $\nu=1,2,3$ and the normalization constant $N_\nu$ is again fixed by $\Phi^L_\nu\Phi^R_\nu=1$.

\section{The appearance of the exceptional points}

\begin{figure}
\includegraphics[width=\textwidth]{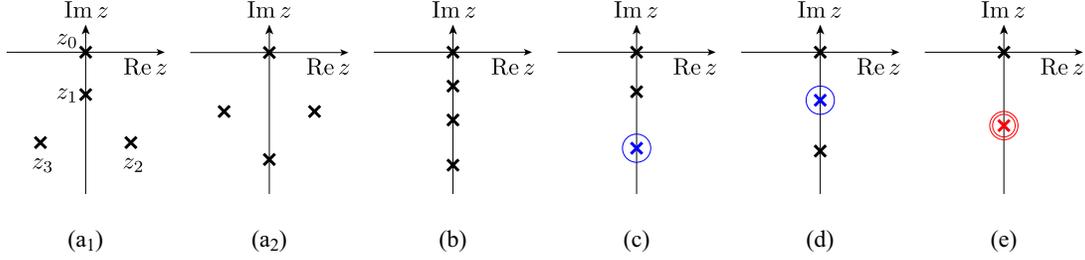}
\caption{Possible distribution of the four eigenvalues. One is always the null eigenvalue.
(a$_1$), (a$_2$) and (b) Possible distributions without any exceptional points.
(c) and (d) Distributions with a second-order exceptional point (blue cross with a single circle).
(e) Distribution with a third-order exceptional point (red cross with a double circle).
}
\label{fig3}
\end{figure}

There are generally two possibilities in the distribution of the eigenvalues, as is illustrated in Fig.~\ref{fig3}~(a$_1$),~(a$_2$) and~(b).
Because of the symmetry that we pointed out above, the eigenvalues are distributed symmetrically with respect to the imaginary axis.

The eigenvalue $z_1$ in Eq.~\eqref{eq220} is always pure imaginary;
this is obvious for $p^3+q^2>0$ because then $u$ and $v$ are both real, but is also true for  $p^3+q^2<0$ because then $u^\ast=v$.
Because of the same symmetry of $u$ and $v$, the two eigenvalues $z_2$ and $z_3$ in Eqs.~\eqref{eq230}--\eqref{eq240} are located symmetrically with respect to the imaginary axis for $p^3+q^2>0$ (Fig.~\ref{fig3}~(a$_1$) and~(a$_2$)), while they are pure imaginary for $p^3+q^2<0$ (Fig.~\ref{fig3}~(b)).

Between Fig.~\ref{fig3}~(a$_1$) and~(b) as well as Fig.~\ref{fig3}~(a$_2$) and~(b), a second-order exceptional point appear as shown in Fig.~\ref{fig3}~(c) and~(d) when the eigenvalues $z_2$ and $z_3$ coalesce on the imaginary axis.
As we discussed at the end of Sec.~\ref{sec1}, 
an exceptional point would normally appear as discrete points in a two-dimensional parameter space. 
In the present case, however, 
because of the symmetry $z_2^\ast=-z_3$ for $p^3+q^2>0$,
the equation $z_2=z_3$ reduces to one condition $\mathop{\mathrm{Re}}z_2=0$ to produce a second-order exceptional point (Fig.~\ref{fig3}~(c) and~(d)).
We therefore have a curve of the second-order exceptional point in the two-dimensional parameter space.
When we further equate $z_2=z_3$ with $z_1$, we have a third-order exceptional point (Fig.~\ref{fig3}~(e)) as a discrete point in the two-dimensional parameter space.

\begin{figure}
\centering
\includegraphics[width=0.5\textwidth]{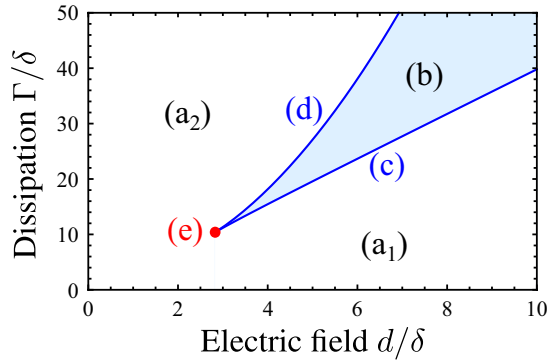}
\caption{The second-order exceptional point appears on the blue curves. 
The third-order exceptional point appears at the end point indicated by a red solid circle.
The eigenvalues are distributed as in Fig.~\ref{fig3}~(b) in the shaded area, while as in Fig.~\ref{fig3}~(a$_1$) or~(a$_2$) outside it.}
\label{fig4}
\end{figure}

The condition $\mathop{\mathrm{Re}}z_2=0$ is equivalent to $u=v$, or $p^3+q^2=0$, which is a quadratic equation for $\Gamma^2$.
Therefore we have two curves of the exceptional points in the range of $\Gamma>0$:
\begin{align}\label{eq320}
\tilde{\Gamma}^{(\pm)}_\mathrm{EP2}
=\sqrt{\frac{\tilde{d}^4}{2}
+10\tilde{d}^2-4
\pm\frac{\tilde{d}}{2}\left(\tilde{d}^2-8\right)^{3/2}
},
\end{align}
where $\tilde{\Gamma}=\Gamma/\delta$ and $\tilde{d}=d/\delta$.
These curves are indicated in Fig.~\ref{fig4}.
The eigenvalue $z_2=z_3$ on the curves is given by
\begin{align}
z_\mathrm{EP2}^{(\pm)}=-\frac{2i\delta}{3}\left[
\tilde{\Gamma}^{(\pm)}_\mathrm{EP2} 
-\frac{1}{4}\sqrt{\frac{\tilde{d}^4}{2}-2\tilde{d}^2-16
\pm\frac{\tilde{d}}{2}\left(\tilde{d}^2-8\right)^{3/2}
}\right],
\end{align}
which is indicated by the blue cross in Fig.~\ref{fig3}~(c) and~(d).


These expressions are only valid when $\tilde{d}\geq 2\sqrt{2}$ and $\tilde{\Gamma}\geq 6\sqrt{3}$, below which the exceptional point goes out of the real parameter space.
The two curves~\eqref{eq320} merge at the point
\begin{align}
\tilde{d}_\mathrm{EP3}=2\sqrt{2}
\quad\mbox{and}\quad
\tilde{\Gamma}_\mathrm{EP3}=6\sqrt{3},
\end{align}
which is indicated by the dot in Fig.~\ref{fig4}.
At this end point, the three eigenvalues coalesce as $z_1=z_2=z_3$ because $u=v=0$, producing a third-order exceptional point, as is illustrated in Fig.~\ref{fig3}~(e).
The eigenvalue at the third-order exceptional point is given by
\begin{align}
z_\mathrm{EP3}=-4\sqrt{3}i\delta.
\end{align}

\section{Summary}

We have presented a non-Hermitian quantum dynamics in which the second-order exceptional point forms curves and a third-order exceptional point appears in a two-dimensional parameter space.
This is due to a symmetry that the non-Hermitian matrix observes.

The system in Fig.~\ref{fig2} may be realizable in experiments.
Admittedly, the exceptional points appear for relatively large values of the environmental coupling $\Gamma$ compared to the detuning parameter $\delta$, for which the Markovian approximation may be questionable.
Let us consider the case in which $\Delta$ is of the order of 1 meV, with which an electric field of frequency 1.5 THz resonates.
We may be able to control the detuning $\delta=\Delta-\omega$, let us say, about 1\% of $\Delta$, which is of the order of 15 GHz, or 10 $\mu$V.
In order to achieve the third-order exceptional point, therefore, the amplitude of the electric field should be of the order of 30 $\mu$V and the environmental coupling $\Gamma$ should be of the order of 100 $\mu$V, if the present argument under the Markovian approximation would be legitimate.

The present argument does not tell anything as to whether the non-Merkovianity is negligible or not in the region $\Gamma/\delta\sim10$, nor, if not, whether the third-order exceptional point in the region survives the non-Markovianity.
Nonetheless, the present demonstration motivates us to look for situations in which the third-order exceptional point exists in a more legitimate range of system parameters, such as systems with more levels.
If such a system is achieved in experiments, it will be a precious example with a third-order exceptional point.
Interesting phenomena to observe include alternation of three eigenstates in an adiabatic encircling of the third-order exceptional point and the state selection in a non-adiabatic encircling.

\section*{Acknowledgements}
The present study was carried out in an exceptionally pleasant collaboration with Prof.~Nimrod Moiseyev when he invited the present author for a splendid stay at his group in Technion.
In the normal situation, he would be on the list of the authors, but it is obviously awkward to have him as a co-author of the paper that the present author dedicates to him.
The present author is also grateful to the generous financial support from Technion.

\bibliographystyle{tfo}
\bibliography{hatano}

\appendix

\section{Derivation of the Lindblad equation with the time-independent Hamiltonian~\eqref{eq110}}
\label{appB}

We here present the derivation of the Lindblad equation with the time-independent Hamiltonian~\eqref{eq110}.
We start with the Lindblad equation
\begin{align}\label{eqb100}
i\frac{d\rho}{dt}=[H,\rho]+i\Gamma\left(c\rho c^\dag-\frac{1}{2}c^\dag c\rho-\frac{1}{2}\rho c^\dag c\right),
\end{align}
where
\begin{align}\label{eqb60}
H&=\begin{pmatrix}
\Delta & de^{-i\omega t}/2 \\
d e^{i\omega t}/2 & 0
\end{pmatrix},
\\
c&=\begin{pmatrix}
0 & 0 \\
1 & 0
\end{pmatrix},
\quad
c^\dag=\begin{pmatrix}
0 & 1 \\
0 & 0
\end{pmatrix}.
\end{align}
Let us transform the density matrix as in
\begin{align}\label{eqb40}
\rho=U(t)\tilde{\rho}U(t)^\dag
\end{align}
with the unitary matrix
\begin{align}
U(t)=\begin{pmatrix}
e^{-i\omega t} & 0 \\
0 & 1
\end{pmatrix}.
\end{align}
By inserting Eq.~\eqref{eqb40} in Eq.~\eqref{eqb100}, 
we have, on the left-hand side, 
\begin{align}\label{eqb180}
i\frac{d}{dt}\left(U(t)\tilde{\rho}U(t)^\dag\right)
&=i\dot{U}\tilde{\rho}U+iU\dot{\tilde{\rho}}U^\dag+iU\tilde{\rho}\dot{U}^\dag
\nonumber\\
&=\begin{pmatrix}
\omega e^{-i\omega t} & 0 \\
0 & 0
\end{pmatrix}
\tilde{\rho}U
+U\left(i\frac{d\tilde{\rho}}{dt}\right)U^\dag
+U\tilde{\rho}\begin{pmatrix}
-\omega e^{i\omega t} & 0 \\
0 & 0
\end{pmatrix}
\end{align}
On the right-hand side of Eq.~\eqref{eqb100}, the first term gives
\begin{align}\label{eqb190}
\left[H,U\tilde{\rho}U^\dag\right]
&=HU\tilde{\rho}U^\dag-U\tilde{\rho}U^\dag H=
UU^\dag H U\tilde{\rho}U^\dag-U\tilde{\rho}U^\dag H UU^\dag
\nonumber\\
&=U\tilde{H}\tilde{\rho}U^\dag-U\tilde{\rho}\tilde{H}U^\dag
=U\left[\tilde{H},\tilde{\rho}\right]U^\dag,
\end{align}
where we introduced a transformed Hamiltonian
\begin{align}
\tilde{H}&=U^\dag H U=
\begin{pmatrix}
e^{i\omega t} & 0 \\
0 & 1 
\end{pmatrix}
\begin{pmatrix}
\Delta & de^{-i\omega t}/2 \\
de^{i\omega t}/2 & 0
\end{pmatrix}
\begin{pmatrix}
e^{-i\omega t} & 0 \\
0 & 1 
\end{pmatrix}
=\begin{pmatrix}
\Delta & d/2 \\
d/2 & 0
\end{pmatrix}.
\end{align}
The second term on the right-hand side of Eq.~\eqref{eqb100} gives
\begin{align}\label{eqb210}
&cU\tilde{\rho}U^\dag c^\dag-\frac{1}{2}c^\dag cU\tilde{\rho}U^\dag-\frac{1}{2}U\tilde{\rho}U^\dag c^\dag c
\nonumber\\
&=U U^\dag c U\tilde{\rho}U^\dag c^\dag UU^\dag
-\frac{1}{2}UU^\dag c^\dag UU^\dag cU\tilde{\rho}U^\dag-\frac{1}{2}U\tilde{\rho}U^\dag c^\dag UU^\dag c UU^\dag
\nonumber\\
&=U\left(\tilde{c}\tilde{\rho}\tilde{c}^\dag -\frac{1}{2}\tilde{c}^\dag\tilde{c}\tilde{\rho}
-\frac{1}{2}\tilde{\rho}\tilde{c}^\dag\tilde{c}\right)U^\dag,
\end{align}
where we introduced the transformed excitation and relaxation operators 
\begin{align}
\tilde{c}^\dag=U^\dag c^\dag U&=\begin{pmatrix}
e^{i\omega t} & 0 \\
0 & 1 
\end{pmatrix}
\begin{pmatrix}
0 & 1 \\
0 & 0
\end{pmatrix}
\begin{pmatrix}
e^{-i\omega t} & 0 \\
0 & 1 
\end{pmatrix}
=e^{i\omega t}
\begin{pmatrix}
0 & 1 \\
0 & 0
\end{pmatrix}
=e^{i\omega t}c^\dag,
\\\label{eqb101}
\tilde{c}=U^\dag c U&=\begin{pmatrix}
e^{i\omega t} & 0 \\
0 & 1 
\end{pmatrix}
\begin{pmatrix}
0 & 0 \\
1 & 0
\end{pmatrix}
\begin{pmatrix}
e^{-i\omega t} & 0 \\
0 & 1 
\end{pmatrix}
=e^{-i\omega t}
\begin{pmatrix}
0 & 0 \\
1 & 0
\end{pmatrix}
=e^{-i\omega t}c.
\end{align}

Summarizing Eqs.~\eqref{eqb180}-\eqref{eqb101}, we have
\begin{align}
i\frac{d\tilde{\rho}}{dt}+U^\dag
\begin{pmatrix}
\omega e^{-i\omega t} & 0 \\
0 & 0
\end{pmatrix}
\tilde{\rho}
-\tilde{\rho}
\begin{pmatrix}
\omega e^{-i\omega t} & 0 \\
0 & 0
\end{pmatrix}
U
&=\left[\tilde{H},\tilde{\rho}\right]
+i\Gamma\left(c\tilde{\rho}c^\dag -\frac{1}{2}c^\dag c\tilde{\rho}
-\frac{1}{2}\tilde{\rho}c^\dag c\right).
\end{align}
The second and third terms on the left-hand side reduce to
\begin{align}
\begin{pmatrix}
\omega & 0 \\
0 & 0
\end{pmatrix}
\tilde{\rho}
-\tilde{\rho}
\begin{pmatrix}
\omega & 0 \\
0 & 0
\end{pmatrix}
=\left[\begin{pmatrix}
\omega & 0 \\
0 & 0
\end{pmatrix},
\tilde{\rho}
\right].
\end{align}
Subtracting this term from the first term on the right-hand side, we have
\begin{align}
i\frac{d\tilde{\rho}}{dt}&=\left[\tilde{H}-\begin{pmatrix}
\omega & 0 \\
0 & 0
\end{pmatrix}
,\tilde{\rho}\right]
+i\Gamma\left(c\tilde{\rho}c^\dag -\frac{1}{2}c^\dag c\tilde{\rho}
-\frac{1}{2}\tilde{\rho}c^\dag c\right).
\end{align}
Leaving out the tilde marks for brevity, we arrive at the updated Lindblad equation of the same form as Eq.~\eqref{eqb100} but with 
the Hamiltonian  now given by the time-independent one
\begin{align}
H=\begin{pmatrix}
\Delta-\omega & d/2 \\
d/2 & 0
\end{pmatrix}=
\begin{pmatrix}
\delta & d/2 \\
d/2 & 0
\end{pmatrix}
\end{align}
instead of the time-dependent rotating-wave approximation~\eqref{eqb60}, where
\begin{align}
\delta=\Delta-\omega=\Ee-\Eg-\omega.
\end{align}

\end{document}